\documentclass[aps,10pt,prd,twocolumn,groupedaddress,nofootinbib,notitlepage,eqsecnum,preprintnumbers]{revtex4-1}

\usepackage[utf8]{inputenc}
\usepackage{hyperref}
\usepackage{xcolor}
\usepackage{graphicx}
\usepackage{amsmath,amssymb,amsbsy,amstext,amsthm}
\usepackage{mathtools}
\usepackage{amsfonts}
\usepackage{comment}
\usepackage{bm}
\usepackage{tabularx}
\usepackage{array,collcell}
\usepackage{enumerate}   

\newcolumntype{C}{>{\collectcell\docellC}c<{\endcollectcell}}
\newcolumntype{L}{>{\collectcell\docellL}c<{\endcollectcell}}
\newcolumntype{R}{>{\collectcell\docellR}c<{\endcollectcell}}

\makeatletter
\providecommand{\@nameedef}[1]{\expandafter\edef\csname#1\endcsname}
\newcommand{\docell}[2]{%
  \sbox\equalizedtablebox{#2}%
  \ifdim\wd\equalizedtablebox>\@nameuse{finallen\theequalizedtable}\relax
    \global\@nameedef{finallen\theequalizedtable}{\the\wd\equalizedtablebox}%
  \fi
  \makebox[\@nameuse{startinglen\theequalizedtable}][#1]{#2}%
}
\newcommand{\docellC}[1]{\docell{c}{#1}}
\newcommand{\docellL}[1]{\docell{l}{#1}}
\newcommand{\docellR}[1]{\docell{r}{#1}}
\newcounter{equalizedtable}
\newsavebox\equalizedtablebox

\makeatother

\begin{document}
\preprint{YITP-20-163}
\title{Approximate gauge independence of the induced gravitational wave spectrum}

\author{\textsc{Guillem Dom\`enech$^{a}$}}
    \email{{domenech}@{pd.infn.it}}
\author{\textsc{Misao Sasaki$^{b,c,d}$} }
    \email{{misao.sasaki}@{ipmu.jp}}

\affiliation{$^{a}$ \small{INFN Sezione di Padova, I-35131 Padova, Italy}\\
      $^{b}$\small{Kavli Institute for the Physics and Mathematics of the Universe (WPI), Chiba 277-8583, Japan}\\
      $^{c}$\small{Center for Gravitational Physics, Yukawa Institute for Theoretical Physics, Kyoto University, Kyoto 606-8502, Japan}\\
      $^{d}$\small{Leung Center for Cosmology and Particle Astrophysics, National Taiwan University, 
      Taipei 10617, Taiwan}
    }

\begin{abstract}
Gravitational waves (GWs) induced by scalar curvature fluctuations are an important source of the cosmological GW background 
and a crucial counterpart of the primordial black hole scenario. However, doubts have been cast on the theoretically predicted induced GW spectrum 
due to its seeming gauge dependence. In this paper, we shed light on the gauge dependence issue of the induced GW spectrum in general cosmological
backgrounds. First, inspired by the Hamiltonian formalism we provide very simple formulas for the tensor modes at second order 
in cosmological perturbation theory. We also emphasize the difference between observable and gauge invariant variables. 
Second, we argue that the Newton (or shear-free) gauge is suitable for both the calculation of induced GWs and the physical interpretation. 
We then show that, most notably, the induced GW spectrum is invariant under a set of reasonable gauge transformations, 
i.e. physically well behaved on small scales, once the source term has become inactive.
 This includes the commonly used flat, constant Hubble and synchronous gauges but excludes the comoving slicing gauge. 
 We also show that a particular solution of the GW equation in a dust dominated universe while the source term is active can be
  gauged away by a small change of gauge. 
\end{abstract}
 \maketitle

\section{Introduction}

In a not so distant future, gravitational wave (GW) detectors such as LISA \cite{Audley:2017drz}, Taiji \cite{Guo:2018npi}, Tianqin \cite{Luo:2015ght},
 DECIGO \cite{Seto:2001qf,Yagi:2011wg}, AION/MAGIS \cite{Badurina:2019hst}, ET \cite{Maggiore:2019uih} 
 and PTA \cite{Lentati:2015qwp,Bian:2020bps} will observe a wide range of frequencies and amplitudes.
  Any detection of a stochastic GW background with a cosmological origin would be a breakthrough in the understanding of the processes of
   the early universe, e.g. see Ref.~\cite{Caprini:2018mtu} for a review.
An important source of cosmological GW backgrounds is the so-called scalar induced GWs \cite{tomita,Matarrese:1992rp,Matarrese:1993zf,Matarrese:1997ay,Carbone:2004iv,Ananda:2006af,Baumann:2007zm,Saito:2008jc}, 
which has received a lot of attention recently \cite{Alabidi:2012ex,Alabidi:2013wtp,Hwang:2017oxa,Espinosa:2018eve,Kohri:2018awv,Cai:2018dig,Bartolo:2018rku,Inomata:2018epa,Yuan:2019udt,Inomata:2019zqy,Inomata:2019ivs,Chen:2019xse,Yuan:2019wwo,DeLuca:2019ufz,Tomikawa:2019tvi,Gong:2019mui,Inomata:2019yww,Yuan:2019fwv,Hwang:2017oxa,Domenech:2017ems,DeLuca:2019ufz,Gong:2019mui,Inomata:2019yww,Yuan:2019fwv,Domenech:2019quo,Ota:2020vfn,Cai:2019jah,Yuan:2019wwo,Cai:2019elf,Cai:2019amo,Bhattacharya:2019bvk,Pi:2020otn}. 
They are a crucial counterpart of the primordial black hole scenario \cite{Espinosa:2018eve,Cai:2018dig,Bartolo:2018rku,Yuan:2019udt} 
and they constitute a powerful probe of the primordial curvature power spectrum \cite{Inomata:2018epa,Gow:2020bzo} 
and of the thermal history of the universe \cite{Cai:2019cdl,Hajkarim:2019nbx,Domenech:2019quo,Domenech:2020kqm}. 
Unfortunately, doubts have been cast on the gauge independence of the derived induced GW spectrum \cite{Hwang:2017oxa}.
 This is an issue rooted in the fact that tensor modes mix with scalar and vector modes at second order in cosmological perturbation theory. 
 It is thus very important to show that there is no ambiguity in the theoretical predictions of the induced GWs.

Since Ref.~\cite{Hwang:2017oxa} first pointed out that the induced GW spectrum is strictly speaking gauge dependent 
(see also Ref~\cite{Tomikawa:2019tvi} for numerical studies in general cosmological backgrounds), there has been a lot of discussion and 
proposed solutions to the gauge issue \cite{Gong:2019mui,Tomikawa:2019tvi,Wang:2019zhj,DeLuca:2019ufz,Inomata:2019yww,Yuan:2019fwv,
	Nakamura:2019zbe,Giovannini:2020qta,Lu:2020diy,Chang:2020tji,Ali:2020sfw,Chang:2020iji,Chang:2020mky,Giovannini:2020soq}. 
Most of the solutions can be classified in either building a gauge invariant formulation of tensor modes at second order \cite{Wang:2019zhj,Yuan:2019fwv,Nakamura:2019zbe,Chang:2020tji,Chang:2020iji,Chang:2020mky} 
or finding an appropriate gauge choice that best describes the GW detection \cite{DeLuca:2019ufz,Inomata:2019yww}. 

The problem with the former approach, i.e., the gauge invariant formulation is that there is no clear connection between 
the observable and the gauge invariant variable. In particular, this approach seems to miss that choosing to work with a particular gauge invariant 
combination is no different than choosing a particular gauge.
The obvious difference is that in the former the degrees of freedom are naturally reduced while in the latter they are reduced by hand. 
In the end, the naive GW spectrum still depends on the choice of a gauge invariant variable and so the gauge issue of Ref.~\cite{Hwang:2017oxa} remains. 

In contrast, although the latter approach taken by Ref.~\cite{DeLuca:2019ufz} certainly goes in the right direction,
defining the observable strain of GWs onto a GW detector  at second order in perturbation theory 
seems to be challenging in a general cosmological background. 
A good argument presented in Ref.~\cite{DeLuca:2019ufz} is that the most suitable gauge is the gauge 
where the coordinates follow a geodesic congruence, e.g. a frame where the mirrors of the interferometer are fixed, 
also known as synchronous gauge.\footnote{In Ref.~\cite{DeLuca:2019ufz} they refer to this gauge as transverse-traceless gauge. 
However, since this only applies to first order in perturbation theory, we find that calling it synchronous gauge is more appropriate in general.} 
However, their reasoning only relies on first order perturbation theory applied to the time delay and does not deal with second order terms. 
Nonetheless, the argument is more convincing when it is shown that the induced GW spectrum in the synchronous gauge exactly matches 
at late times with the induced GW spectrum in the Newton (or shear-free) \cite{DeLuca:2019ufz,Inomata:2019yww}. 
Yet, in the absence of a well-defined observable which is invariant under second order gauge transformations, the gauge issue persists. 
Moreover, as we shall see, the synchronous gauge is a subtle gauge for the induced GWs due to its remaining gauge degrees of freedom, as was first noted in Ref.~\cite{Lu:2020diy}. 

A third solution would be to build a GW energy momentum tensor which is gauge invariant under second order gauge transformations. 
However, although Isaacson \cite{Isaacson:1967zz,Isaacson:1968zza}  showed that one can find a well-defined 
energy momentum tensor for GWs in the limit of short wavelengths, i.e. short compared to the curvature of spacetime, in the case of vacuum 
(Ricci flat) spacetime, there is yet no such object in a cosmological background. 
This stems from the fact that in General Relativity there is no well-defined notion of localized energy. 
A study in this direction was done in \cite{Mukhanov:1996ak}, but only for the first order perturbation quantities, in the context of
the backreaction of perturbations to the background cosmological evolution.
In our case of interest, for instance, one may include terms quartic in gradients of scalar quantities in the energy density of GWs. 
In the end, these additional terms at fourth order may render the total GW spectrum gauge invariant but this direction might actually
 miss the essence of the current discussion.
In similar lines, Ref.~\cite{Giovannini:2020qta,Giovannini:2020soq} argued that second order gauge transformations yield 
a spurious contribution to the observed GW spectrum and proposed some candidates for the definition of the GW spectral density. 
Furthermore, Ref.~\cite{Giovannini:2020soq} reasons that the induced GW spectrum is only meaningful in completely fixed gauges like 
the Newton or uniform curvature gauges. This conclusion is in contrast with Ref.~\cite{DeLuca:2019ufz}.

In this paper, we explore a new direction which we believe reduces the gauge issue of the induced GW spectrum to a minimum. 
By analogy with GWs generated by mergers of binary black holes, where GWs are only well defined far enough from the source, 
we argue that in cosmology the energy density of GWs for modes deep inside the horizon is well defined and gauge invariant 
as long as: ($i$) the source of induced GWs is not active and ($ii$) the spacetime slicing is well behaved on small scales. 
This implies that the induced GW spectrum is invariant under a set of reasonable gauge transformations on subhorizon scales. 
We show such approximate gauge invariance for a general cosmological background filled with a perfect fluid with constant 
equation of state and constant speed of sound, $p/\rho=w=c_s^2=\rm constant$.
We also present very simple formulas for both the source term of induced GWs and the transformation of tensor 
modes at second order in a general gauge, which helps to clarify the gauge issue.
 In this way, our discussion is not obscured by the involved calculations at second order perturbation theory.

The paper is organized as follows. In Sec.~\ref{sec:gaugeinvariance} we present simple formulas for the source term of induced GWs 
in a general gauge inspired by the Hamiltonian formalism developed in Ref.~\cite{Domenech:2017ems}. 
We also emphasize the difference between the gauge invariant formalism, a gauge choice and the observable energy density. 
In Sec.~\ref{sec:GWspectrum}, we define a class of physically meaningful gauges on small scales and show that the induced GW spectrum is invariant 
under such a set of gauges on subhorizon scales. We pay special attention to the synchronous gauge. In Sec.~\ref{sec:dustdom} we argue that 
the induced GWs generated in pressure-less adiabatic perfect fluid dominated universe can be gauged away at the stage when the source term is active. 
Lastly, in Sec.~\ref{sec:conclusions} we discuss the implications of our work. 
Details of the calculations and gauge transformations can be found in the appendices.
 Throughout the paper we work in reduced Planck units where $c=\hbar=8\pi G=1$.

\section{Gauge invariance vs observable \label{sec:gaugeinvariance}}

In this section, we focus on the difference between a gauge invariant quantity and an observable quantity. 
We also derive simple formulas for the generation of tensor modes by scalar squared terms at second order perturbation theory 
and briefly review the gauge issue raised in Ref.~\cite{Hwang:2017oxa}.

We begin by describing the universe content and our perturbative expansion which closely follows that of Ref.~\cite{Domenech:2017ems}. 
We consider that the universe is filled with a perfect fluid with energy-momentum tensor given by
\begin{align}
T_{\mu\nu}&=(\rho+p)u_{\mu}u_{\nu}+p g_{\mu\nu}\,,
\end{align}
where $\rho$ and $p$ respectively are the energy density and pressure and $u_{\mu}$ is the fluid 4-velocity. 
We take the spacetime metric to be a perturbed flat Friedmann-Lema{\^i}tre-Robertson-Walker (FRLW) universe.
 By means of the (3+1)-decomposition we may write the metric as
\begin{align}\label{eq:confdecom}
ds^2&=a^2(\tau)\nonumber\\&\times\left(-N^2d\tau^2+{\rm e}^{2\phi}\Upsilon_{ij}\left(dx^i+N^idt\right)\left(dx^j+N^jdt\right)\right)\,,
\end{align}
where $a$ is the scale factor, $N$ is the lapse, $N^i$ is the shift vector and $\phi$ and $\Upsilon_{ij}$ respectively represent 
the trace and traceless degrees of freedom of the spatial metric. 
We perturbatively expand the lapse, shift and $\Upsilon_{ij}$ as
\begin{align}\label{eq:confdecom2}
N&=1+\alpha\quad,\quad N_i=\partial_i\beta\\
\left[\ln\Upsilon\right]_{ij}&=h_{ij}+2\left(\partial_{i}\partial_{j}-\tfrac{1}{3}\delta_{ij}\Delta\right)E\label{eq:confdecomh}\,,
\end{align}
where  $\delta^{ij}h_{ij}=\partial^i h_{ij}=0$ and we have neglected vector modes for simplicity. Thus $h_{ij}$  represents the transverse-traceless degrees of freedom, i.e.
 the tensor degrees of freedom. We note, however, that this does not necessarily mean $h_{ij}$ defined here is a directly observable quantity.
As we shall see, the exponential form of the spatial metric in Eq.~\eqref{eq:confdecom} simplifies substantially the calculations and discussions. 
For more details on the perturbative expansion see Ref.~\cite{Domenech:2017ems}. 

It is important to observe that the relation between 
the metric perturbations in our expansion \eqref{eq:confdecom} and the commonly used variables \eqref{eq:confdecom3}, 
e.g. in Refs.~\cite{Malik:2008im,Hwang:2017oxa,Gong:2019mui}, involve a mixture of scalar and tensor modes, see App.~\ref{app:relation}. 
For instance, the tensor modes of our metric \eqref{eq:confdecom} are related to the tensor modes in the common 
expansion \cite{Malik:2008im,Hwang:2017oxa,Gong:2019mui} by additional scalar squared terms containing $E$ and $\phi$. 
Surely, the resulting induced GW spectrum should not depend on the choice of the perturbative expansion. 
Nevertheless, we start to see how the gauge choice may naively affect the spectrum of induced GWs.

At second order in cosmological perturbation theory, the squared of gradients of scalar variables source tensor modes. 
In the perturbative expansion \eqref{eq:confdecom}, the equations of motion for the tensor modes at second order without 
any gauge fixing take a very simple form, concretely
\begin{align}\label{eq:heom}
(\hat D_\tau-\Delta) h_{ij}=\widehat{TT}^{ab}_{ij}S_{ab}\,,
\end{align}
where $\hat D_\tau\equiv a^{-2}\partial_\tau(a^2\partial_\tau)$, $\Delta\equiv\delta^{ij}\partial_i\partial_j$ is the flat 3-dimentional Laplacian, 
$\widehat{TT}^{ab}_{ij}$ is the transverse-traceless projector, e.g. found in Ref.~\cite{Domenech:2017ems}, and
\begin{align}\label{eq:heomsource}
S_{ab}=4&\partial_a\Phi\partial_b\Phi+2a^2\left(\rho+p\right)\partial_aV\partial_bV\nonumber\\&-(\hat D_\tau-\Delta)\left[\partial_a\sigma\partial_b\sigma+\partial_a\partial^kE\partial_k\partial_bE\right]\,.
\end{align}
In the source term \eqref{eq:heomsource} we have defined
\begin{align}\label{eq:definitions}
\Phi&\equiv\phi-\tfrac{1}{3}\Delta E+\sigma\quad,\quad V\equiv v+E'\,,\\
\sigma&\equiv\beta-E'\,.
\end{align}
 The variable $\sigma$ corresponds to the scalar component of the shear of the hypersurface,
 $v$ is the scalar component of the fluid 3-velocity and $\Phi$ and $V$ are two different gauge invariant variables under first order gauge transformations, 
 see App.~\ref{app:gauge}. They respectively represent the curvature perturbation and the fluid velocity potential on Newton (or shear-free) slices. 
Equation~\eqref{eq:heomsource} is obtained by means of the Hamiltonian formalism presented in Ref.~\cite{Domenech:2017ems}.
We cross-checked it by the direct expansion of the Einstein equations. To achieve such a simplified form of Eq.~\eqref{eq:heomsource} 
we used several times the first order equations of motion which can be found in App.~\ref{app:eom}. 
Also, we find that the terms containing the variable $E$ in Eq.~\eqref{eq:heomsource} are absent in the general formulas derived in Ref.~\cite{Gong:2019mui}
 and later used in Refs.~\cite{DeLuca:2019ufz,Inomata:2019yww}. This is in agreement with Ref.~\cite{Lu:2020diy}. 
Although the difference is a redefinition of the tensor modes, these terms play a crucial role in the discussion 
in the synchronous gauge of Sec.~\ref{sec:GWspectrum}. 

From Eq.~\eqref{eq:heomsource} we have several insights into the issue at hand. First, we see that the terms in Eq.~\eqref{eq:heomsource} 
which are not gauge invariant at second order can be reabsorbed by a redefinition of the tensor modes. Explicitly by defining
\begin{align}\label{eq:hN}
 h^{N}_{ij}\equiv h_{ij}+\widehat{TT}^{ab}_{ij}\left[\partial_a\sigma\partial_b\sigma+\partial_a\partial^kE\partial_k\partial_bE\right]\,,
\end{align}
we find that
\begin{align}\label{eq:heomN}
(\hat D_\tau-\Delta)& h^{N}_{ij}=\widehat{TT}^{ab}_{ij}\left[4\partial_a\Phi\partial_b\Phi+2a^2\left(\rho+p\right)\partial_aV\partial_bV\right]\,,
\end{align}
has now a gauge invariant form. In Eq.~\eqref{eq:hN} we deliberately used the superscript $N$ to indicate that $h^N_{ij}$ coincides 
with the tensor modes $h_{ij}$ evaluated in the Newton gauge \cite{Domenech:2017ems} where $\sigma=E=0$. 

It should be noted that even though Eq.~\eqref{eq:heomN} is gauge invariant, one may have chosen any other gauge invariant definition 
of the tensor modes. Then, the source term in Eq.~\eqref{eq:heomN} would of course be different and so would be the derived induced GW spectrum. 
This point let us move forward to discuss what is actually the observable GWs. A laser interferometer or an array of pulsars detect the GW strain. 
However, for stochastic GW backgrounds in cosmology one uses the spectral density of GWs defined by
\begin{align}\label{eq:spectraldensity}
\rho_{\rm GW}(k)=\frac{k^3}{16\pi^2}\sum_\lambda\langle h_{\bm{k},\lambda}'h_{-\bm{k},\lambda}'+ k^2 h_{\bm{k},\lambda}h_{-\bm{k},\lambda}\rangle\,,
\end{align}
where $\langle\cdots\rangle$ denotes ensemble average.
For simplicity, we work with the spectral density from now on but our conclusions should not depend on whether one uses the GW strain 
or the GW spectral density. Now, it is obvious that the GWs spectral density as defined in Eq.~\eqref{eq:spectraldensity} is a gauge dependent quantity
if one considers second order gauge transformations. This is solely because $h_{ij}$ changes under a second order gauge transformation. 
In other words, the \textit{naive} GW spectrum obtained using Eq.~\eqref{eq:spectraldensity} of the induced GWs, e.g., in the Newton gauge 
is a priori different from that in the comoving gauge. This does not mean that the GW spectrum is gauge dependent, 
it simply means that energy density of GW is in general not well defined by Eq.~\eqref{eq:spectraldensity} when one considers higher order terms.

The crucial point is to realize that Eq.~\eqref{eq:spectraldensity} should be well-defined on subhorizon scales, $k\gg{\cal H}$, 
as long as the source term of induced GWs in Eq.~\eqref{eq:heomsource} is no longer active. In the absence of a source,
 the tensor modes $h_{ij}$ deep inside the horizon \textit{are} freely propagating GWs. However, there is a catch. 
 For practical convenience, our GW detector should be described in a reasonable coordinate system. 
 Otherwise, if we were in a coordinate system where the GW detector would oscillate wildly, we would most likely confuse GWs with gauge artifacts. 
 In the next section we explore what would be a set of reasonable gauges and how the energy density is invariant under such a set of gauge transformations.

\section{Gauge invariance of GW spectrum \label{sec:GWspectrum}}

In the previous section we showed that the spectrum of induced GW is always gauge dependent if one uses the definition of the spectral density as in Eq.~\eqref{eq:spectraldensity}. However, one expects that if the gauge transformation leaves the subhorizon physics essentially untouched, 
the prediction for the induced GWs should agree independently of the gauge. 
In this sense, the Newton gauge seems suitable for physical interpretations since
one recovers Newtonian gravity for $k\gg{\cal H}$. In this paper, we assume this is indeed the case. In regard to the observable strain, 
we show that $h_{ij}^N$ coincides with that in a suitable fixed synchronous gauge.

Now, let us take a look at the gauge transformation of the Newton potential since we are working with the gauge invariant variables  $\Phi$ 
and $h_{ij}^N$ that coincide with the Newton potential and the tensor modes in the shear-free gauge. As we already argued, 
the Newton gauge is a well-behaved gauge at the smallest scales. Then, we find that the relation between
 the trace part of the metric perturbation $\phi$ evaluated in an arbitrary gauge and in the Newton gauge is given by
\begin{align}\label{eq:phiG}
\Phi_G&= \Phi_N+{\cal H}T_G+\tfrac{1}{3}\Delta L_G\,,
\end{align}
where ${\cal H}\equiv a'/a$, $T_G$ and $L_G$ respectively are the time and spatial gauge parameters 
and we used the subscript $G$ to emphasize that it is for an arbitrary gauge. 
We have also explicitly added the subscript $N$ to $\Phi$, and set $\phi=\Phi_G$ to emphasize that $\Phi_N$ is $\phi$ defined in the Newton gauge
and $\Phi_G$ is $\phi$  in another fixed gauge.
See App.~\ref{app:gauge} for the details on the gauge transformations and App.~\ref{app:eom} for the basic equations of motion. 

We define the requirement that the gauge is well-behaved on subhorizon scales as the condition that 
\begin{align}\label{eq:condition1}
\Phi_G(k\gg{\cal H})=O\left(\Phi_N(k\gg{\cal H})\right)\,.
\end{align}
This implies that the well-behaved set of gauge transformation has the gauge parameters ${\cal H}T_G(k\gg{\cal H})$ 
and $\Delta L_G(k\gg{\cal H})$ which decay equal or faster than $\Phi_N(k\gg{\cal H})$. 
As we shall show, this includes the flat, constant Hubble and synchronous gauges. The requirement that the density contrast $\delta\rho_G(k\gg{\cal H})=O\left(\delta\rho_N(k\gg{\cal H})\right)$ leads 
to a similar conclusion on $T_G$.

The Newton potential in a general cosmological background with a constant equation of state parameter $p/\rho=w=c_s^2$ decays inside 
the horizon roughly as
\begin{align}\label{eq:gravitationalpotential2}
\Phi_N(c_sx\gg 1)\propto  \Phi(k)\, (c_sx)^{-2-b}\,; \quad c_s\neq0\,,
\end{align}
where $\Phi(k)$ is a $k$-dependent amplitude to be set by the initial condition, $x$ and $b$ are defined by
\begin{align}
x\equiv k\tau\quad,\quad b\equiv\frac{1-3w}{1+3w}\,,
\end{align}
and we have neglected the oscillatory behavior inside the sound horizon.
It should be noted that in the special case of a dust universe when $c_s^2=w=0$ the Newton potential is constant in all scales. 
For this very particular limit, the source term of secondary GWs \eqref{eq:heomN} is indefinitely active and our criterion 
for the approximate gauge invariance of induced GWs does not apply.
 In this special case, we show in Sec.~\ref{sec:dustdom} that the induced GWs can be essentially gauged away. 
 This means that during the dust domination one cannot straightforwardly tell apart what is a gauge artifact and what are GWs until 
 the dust domination ends and the universe transitions to an era with $c^2_s\neq0$. 

 We note that we do not claim that the time-independent $h_{ij}$ generated during dust domination is meaningless. 
 We only claim that it can be gauged away as long as the universe is dust dominated.
 It becomes a genuine gravitational wave when the dust domination ends. In a sense, this is similar to the curvature perturbation
 on superhorizon scales during inflation. It can be gauged away as long as the universe is in the inflationary phase, or in an eternally inflating universe. 
It becomes physically significant only after the universe enters a decelerated phase.
 This highlights the particularities of the induced GW generated in a dust dominated universe.
 Note that in models with an early matter dominated stage, the dominant contribution to the induced GW spectrum 
comes mostly from the stage right after reheating \cite{Inomata:2019zqy,Inomata:2019ivs}. 
Our criterion for the approximate gauge invariance applies to this contribution.
 Hence there is no gauge issue with the induced GWs in a universe with an early dust dominated stage
 as long as one considers the stage after dust domination.

We shall translate the reasonable gauge condition \eqref{eq:condition1} into conditions on the time dependence of the gauge parameters as
\begin{align}
T_G(x\gg1)&\propto k^{-1} (c_sx)^{-c_T}\,;\quad c_T\geq1+b\,,
\nonumber\\
L_G(x\gg1)&\propto k^{-2} (c_sx)^{-c_L}\,;\quad c_L\geq2+b\,.
\label{eq:condition2}
\end{align}
Let us show that the conditions on $c_T$ and $c_L$ are enough to ensure the gauge independence of the GW spectrum on subhorizon scales. 
First, we find that the tensor modes in an arbitrary gauge are related to those in the Newton gauge by 
\begin{align}\label{eq:gaugehij}
h^{G}_{ij}=h^{N}_{ij}-\widehat{TT}_{ij}\,^{ab}&\Big\{\partial_aT_G\partial_bT_G+\partial_a\partial_kL_G\partial_b\partial_kL_G\Big\}\,.
\end{align}
Second, the GW spectral density is related to the dimensionless GW spectrum\footnote{The dimensionless power spectrum of a quantity Q is defined by
\begin{align}
    \langle Q(k)Q(k')\rangle=\frac{2\pi^2}{k^3} {\cal P}_{Q}(k)\delta(k+k')\,.
  \end{align}
} by 
\begin{align}\label{eq:OMGWC}
\Omega_{\rm GW}(k\tau\gg1)=\frac{k^2}{12{\cal H}^2}\overline{{\cal P}_h(k\tau\gg1)}\,,
\end{align}
where we used that deep inside the horizon after GWs are generated $h'_{\bm{k}}\sim kh_{\bm{k}}$ and an overline denotes oscillation average. 
When the source of the induced GWs is no longer active, we can treat $h_{ij}$ and $\Phi$ as independent variables. 
In this situation, by using Eq.~\eqref{eq:gaugehij} the GW spectrum in an arbitrary gauge and in the Newton gauge are related by
\begin{align}\label{eq:PGPG}
\overline{{\cal P}^{G}_h(k\tau\gg1)}=\overline{{\cal P}^{N}_h(k\tau\gg1)}+\overline{{\cal P}_G(k\tau\gg1)}\,,
\end{align}
where $\overline{{\cal P}^{N}_h(k\tau\gg1)}$ is presented in App.~\ref{app:eom} and 
\begin{align}\label{eq:PG}
{\cal P}_G&(k\tau\gg1)=\sum_\lambda\frac{k^3}{\pi^2}\int \frac{d^3q}{(2\pi)^3} \,\left(e_\lambda^{ij}(\bm{k})q_iq_j\right)^2 
\nonumber\\
&\times\big|T_{\bm{k}}T_{|\bm{k}-\bm{q}|}+(q^2-k^lq_l)L_{\bm{k}}L_{|\bm{k}-\bm{q}|}\big|^2\,.
\end{align}
In Eq.~\eqref{eq:PG}, $\lambda$ is the GW polarization, $e_\lambda^{ij}(\bm{k})$ is the polarization tensor of the GWs 
that satisfies $\delta_{ij}e_\lambda^{ij}(\bm{k})=k_ie_\lambda^{ij}(\bm{k})=0$
and $e_\lambda^{ij}(\bm{k})e_{ij,\lambda'}(\bm{k})=\delta_{\lambda\lambda'}$.
For our purposes it is enough to notice that $\overline{{\cal P}_G(k\tau\gg1)}$ is quartic in $T_G$ and $L_G$. Note that in the expressions \eqref{eq:PGPG} and \eqref{eq:PG} the assumption of a finite time source translates into a spectrum of scalar fluctuations with a finite width. This means that there is a moment in time when all the scalar fluctuations are inside the horizon and no more secondary GWs are generated. Thus, the integral over momenta in Eq.~\eqref{eq:PG} has finite and non-zero integration limits and we shall use the subhorizon approximation for $T_G$ and $L_G$ for all scales within the integration limits.

For an arbitrary constant equation of state $w$, the analytical formulas for the spectrum of induced GWs $\overline{{\cal P}^{N}_h(k\tau\gg1)}$
 are provided in Ref.~\cite{Domenech:2019quo}. 
The only relevant point for our discussion is the fact that the time dependence of the GW power spectrum is given by
\begin{align}\label{eq:powerh}
\overline{{\cal P}^{N}_h(x\gg1)}\approx x^{-2(1+b)}{\cal P}(k)\,,
\end{align}
where ${\cal P}(k)$ is some $k$ dependent amplitude. We refer the reader to Ref.~\cite{Domenech:2019quo} or App.~\ref{app:eom} for the details. 
In other words, induced GWs inside the horizon behave as a free massless mode or as free GWs, i.e. they oscillate and decay as 
\begin{align}
h_{ij}\propto 1/a(\tau)\propto x^{-1-b}\,.
\end{align}

The condition that the GW spectrum is gauge invariant may be expressed as
 $\overline{{\cal P}^{G}_h(k\tau\gg1)}=\overline{{\cal P}^{N}_h(k\tau\gg1)}$. This condition implies from Eq.~\eqref{eq:condition2} that
\begin{align}\label{eq:condition3}
c_T\,,\,c_L\geq (1+b)/2\,.
\end{align}
The conditions \eqref{eq:condition3} are always satisfied as long as conditions \eqref{eq:condition2} are satisfied. 
Thus, our requirement of a well-behaved gauge on small scales yields that the GW spectrum is invariant under such a reasonable set of
 gauge transformations. Note that this conclusion is independent on whether one uses the GW strain or the GW spectral density. 
We shall proceed to show for which commonly used gauges conditions Eq.~\eqref{eq:condition2} are satisfied.

For simplicity let us start with a change of the time slicing. This means we can set $L_G=0$ in our gauge transformation. 
Thus, all the discussion lies on whether the gauge parameter $T_G$ satisfies the conditions \eqref{eq:condition2} or not. 
In App.~\ref{app:transformationrules} we present the detailed gauge transformation rules. 

Let us start with a bad example. 
We consider the comoving slicing gauge where the fluid 4-velocity is orthogonal to the constant time hypersurfaces,
or the hypersurface normal vector agrees with the rest frame of the fluid. This slicing is fixed by setting $v+\beta=0$. 
We find that at late times and on subhorizon scales, the temporal gauge parameter in the comoving slicing reads
\begin{align}\label{eq:TC}
T_C(x\gg1)\propto\frac{\Phi(k)}{c_sk}(c_sx)^{-b}\,,
\end{align}
where the subscript $C$ stands for comoving. From Eq.~\eqref{eq:TC} it is clear that the comoving gauge does not satisfy 
the criteria \eqref{eq:condition2} independently of $w$ (or $b$). This means that the induced GW spectrum evaluated 
using \eqref{eq:OMGWC} is quite different in the comoving slicing gauge compared to the Newton gauge. 
This is not so surprising since on short distances one expects that the fluid density and velocity oscillate. 
Thus the requirement that the slicing  is comoving with the fluid would imply a highly deformed slicing in the small scale limit where
the spacetime is almost flat, and would most likely give rise to gauge artifacts in the GW spectrum.
We conclude that the comoving slicing is not well-behaved on small scales 
and extra caution should be applied if the induced GW spectrum is computed in such a gauge.

Let us turn to two good examples. The first example is the spatially flat slicing where the curvature of the intrinsic metric (or the spatial curvature) is
homogeneous. This is achieved by setting $\phi=E=0$. The temporal gauge parameter to go to the flat gauge from the Newton gauge is
\begin{align}\label{eq:TF}
T_F(x\gg1)\propto\frac{\Phi(k)}{c_sk}\, (c_sx)^{-1-b}\,,
\end{align}
where $F$ stands for flat slicing.
The second example is the constant Hubble slicing, where the extrinsic curvature is homogeneous. 
This implies that $3\phi'-3{\cal H}\alpha-\Delta\beta=0$. 
The time gauge parameter from the Newton gauge to the constant Hubble gauge reads
\begin{align}\label{eq:TH}
T_H(x\gg1)\approx\frac{c_s\Phi(k)}{k}(c_sx)^{-2-b}\,,
\end{align}
where $H$ stands for constant Hubble slicing. 
We see that Eqs.~\eqref{eq:TF} and \eqref{eq:TH} satisfy the conditions \eqref{eq:condition2} for a reasonable gauge transformation. 
This was expected since deep inside the horizon the cosmology should be irrelevant, e.g. the expansion rate or the spatial curvature. 
We conclude that the induced GW spectrum computed in the Newton, flat and constant Hubble gauge coincide for scales deep inside
 the horizon on an arbitrary background. Thus we have shown the approximate gauge invariance of the induced GW spectrum.

\subsection{Synchronous gauge}

We dedicate a separate subsection for the synchronous gauge as the analysis is more subtle. This is because the synchronous gauge in which 
one sets $\alpha=\beta=0$ does not completely kill the gauge degrees of freedom.
There remains a residual gauge ambiguity in the variables. For example see Ref.~\cite{Malik:2008im} or App.~\ref{app:transformationrules}. 
As we shall see, it turns out that this ambiguity is very much related to the terms in the equations of motion of secondary GWs \eqref{eq:heom} 
which contain the variable $E$ and that have been neglected in previous analyses \cite{DeLuca:2019ufz,Inomata:2019yww}.

We find that the gauge parameters that relate the synchronous gauge with the Newton gauge are given by
\begin{align}\label{eq:TS}
T_S(x\gg1)&\approx \frac{x^{-1-b}}{c_s k}\left(-{\Phi(k)}+ \tilde T_0(k)\right)+O(x^{-2-b})\,,\\
L_S(x\gg1)&\approx \frac{x^{-b}}{c_s^2 k^2}\left({\Phi(k)}-\tilde T_0(k)\right)+O(x^{-2-b})\,,\label{eq:LS}
\end{align}
where $\tilde T_0(k)$ is an arbitrary function of the wavenumber $k$ and reflects the residual gauge ambiguity in the synchronous gauge. 
For simplicity, in Eq.~\eqref{eq:LS} we have fixed the residual gauge ambiguity in the spatial gauge by eliminating a constant after integration. 
See App.~\ref{app:transformationrules} for more details. We see that while the temporal gauge parameter $T_S$ \eqref{eq:TS} satisfies 
conditions \eqref{eq:condition2}, the spatial gauge parameter $L_S$ \eqref{eq:LS} does not in general. Nevertheless, we see that $L_S$ \eqref{eq:LS} satisfies the criterion of a well-behaved gauge if one properly fixes $\tilde T_0(k)$ 
to remove the leading order terms on the right hand side of Eqs.~\eqref{eq:TS} and \eqref{eq:LS}. 
Thus, we conclude that a properly fixed synchronous gauge yields the same induced GW spectrum as the one in the Newton gauge. 
Nevertheless, the above discussion shows that the synchronous gauge might not be a suitable gauge for the computation of 
induced GWs due to the residual gauge ambiguity, in contrast to the argument of Ref.~\cite{DeLuca:2019ufz}. 

At this point, let us compare our results with those of Refs.~\cite{DeLuca:2019ufz,Inomata:2019yww}. As is clear from Eq.~\eqref{eq:heomsource}, the source term of the induced GWs in the synchronous gauge contains terms proportional to $E$. However, in Refs.~\cite{DeLuca:2019ufz,Inomata:2019yww} it seems these terms have been neglected. We show in App.~\ref{app:relation} that in the notation of Refs.~\cite{DeLuca:2019ufz,Inomata:2019yww} additional terms proportional to $E$ also appear, see Eq.~\eqref{eq:heomsource2}. In the end though, our conclusions agree with Refs.~\cite{DeLuca:2019ufz,Inomata:2019yww} because these neglected terms proportional to $E$ are in fact the ones that make a difference in the gauge transformation. 

Now, it is important to note that if one solves the equations of motion for the induced GWs \eqref{eq:heomsource} in the synchronous gauge  by requiring that the variables $\sigma$ and $E$ are well-behaved on superhorizon scales as in Refs.~\cite{DeLuca:2019ufz,Inomata:2019yww}, one would end up with an induced GW spectrum different compared to the one in the Newtonian gauge. Nevertheless, as we showed in this section, the difference is clearly a gauge artifact as one can do at any time a gauge transformation within the synchronous gauge that removes the spurious contribution. Thus, in the end, one finds that the induced GW spectrum of a properly fixed synchronous gauge agrees with that computed in the Newtonian gauge.

Before ending this section, we note that although the synchronous gauge is most 
convenient for computing the response of a gravitational wave detector, one can 
of course choose a different gauge without affecting the physical result, provided 
that one carefully performs the calculation. In fact, when computing 
the detector response, one has to introduce a frame in which the detector is at rest. 
This can be done in any gauge, but the synchronous gauge is convenient since the
coordinates can be chosen such that the time coordinate coincides with the proper
time of the rest frame and the spatial coordinates are comoving with the detector.
This also means that the synchronous gauge is not always convenient unless
the remaining gauge degrees of freedom are adequately fixed. If not, one may
encounter a spurious component in the metric that could lead to a wrong result.
For example, if we work in a coordinate system spanned by a family of geodesics that 
does not match the rest frame of the detector, one could obtain a singular result 
if those geodesics happen to have focusing singularities in the vicinity of the 
worldline of the detector.

\section{The dust dominated universe\label{sec:dustdom}}
In this section, we argue that GWs induced during a dust dominated universe, i.e. $w=c_s^2=0$, can be gauged away as long as the source term persists.
There is in fact a reasonable suspicion. During dust domination the Newton potential is constant in time on all scales and, 
therefore, so is the source term in Eq.~\eqref{eq:heomN}. A constant source to the tensor modes leads to a constant particular solution to $h_{ij}$ 
which does not behave like a GW by any means. To show that it is indeed a gauge artifact let us demonstrate that there 
is a particular gauge transformation by which the solution essentially vanishes.

We start from the gauge invariant form of the equations of motion \eqref{eq:heomN}. The solutions to $\Phi_N$ and $V_N$
 in a dust universe are given in App.~\ref{app:eom}. Here we use the first order equations of motion to replace 
 the gauge invariant variable $V_N$ in favour of $\Phi_N$. In this way, we obtain
\begin{align}\label{eq:heomMD}
(\hat D_\tau-\Delta)h^N_{ij}=\widehat{TT}^{ab}_{ij}\left\{\frac{20}{3}\partial_a\Phi_N\partial_b\Phi_N\right\}\,.
\end{align}
Since $\Phi_N$ is constant in time there is a particular solution in which $h_{ij}$ is constant in time. 
In fact in Fourier space, the solution after requiring that $h_k^N|_{\tau=0}=h_k^{N'}|_{\tau=0}=0$ reads \cite{Mollerach:2003nq,Hwang:2017oxa}
\begin{align}\label{eq:hMD}
h^N_{\bm{k}}(x)=k^{-2}S^N_{k}\left(1-3j_1(x)/x\right)\,,
\end{align}
where $j_1(x)$ is the spherical Bessel of the first kind of order one and
\begin{align}\label{eq:SMD}
S^N_k=\frac{20}{3}\int \frac{d^3q}{(2\pi)^3}e^{ij}(\bm{k})q_iq_j\Phi_{\bm{q}}\Phi_{\bm{k}-\bm{q}}\,.
\end{align}
As is clear from Eq.~\eqref{eq:hMD}, the tensor modes are initially zero at $x\ll1$ and 
develop a constant value $h_{\bm{k}}\sim k^{-2}S_{k}$ on subhorizon scales, that is at $x\gg1$. 
Before we show that this constant solution is likely to be a gauge artifact it is important to note that the induced tensor modes \eqref{eq:hMD} 
do not behave as what is usually considered to be GWs in an expanding universe.
First, its amplitude does not decay proportional to $1/a$ and, therefore, 
the energy density of such GWs does not decay as $1/a^{4}$, as is expected from a fluid made of massless fields such as photons. 
Second, in the flat spacetime limit, the observable effect would be a time-dependent strain \cite{Maggiore:1900zz,Flanagan:2005yc}.
This is clear from noting that the geodesic deviation equation is proportional to $h''_{ij}$. Therefore, a GW detector in a dust universe would not 
detect the strain of \eqref{eq:hMD} as a GW. 
So instead of GWs, one should refer to the induced tensor mode \eqref{eq:hMD} as a static anisotropic stress. This was noted in Refs.~\cite{Assadullahi:2009nf,Inomata:2019yww}.

We proceed to argue that the induced tensor modes \eqref{eq:hMD} are gauge artifacts. To do that, we look for a suitable gauge transformation 
which brings the amplitude of \eqref{eq:hMD} to zero. Since a static $h_{ij}$ implies a non-vashing intrinsic Riemann tensor of the spatial hypersurfaces,
it can't be gauged away by spatial gauge transformation. Therefore we set $L_G=0$ and 
focus only on a temporal gauge transformation with $T_G$ constant in time. In this case, we have that
\begin{align}\label{eq:heomMD2}
h^G_{\bm{k},\lambda}=h^N_{\bm{k},\lambda}
-\int \frac{d^3q}{(2\pi)^3}e^{ij}(\bm{k})q_iq_jT_{G,\bm{q}}T_{G,\bm{k}-\bm{q}}\,,
\end{align}
where $h^N_{\bm{k}}$ is given by the constant term of Eq.~\eqref{eq:hMD}. 
We shall present an example of $T_G$ which cancels the contribution from $h^N_{\bm{k}}$, which is only a function of $k$, 
up to an arbitrary small contribution. We choose the Fourier transform of $T_G$ to be of the form
\begin{align}\label{eq:TGMD}
T_{G,\bm{q}}=\frac{(2\pi)^{3/4}F(q)}{\varepsilon^{3/2}}
\left(e^{-\frac{(\bm{q}-\bm{q}_+)^2}{2\varepsilon^2}}+e^{-\frac{(\bm{q}-\bm{q}_-)^2}{2\varepsilon^2}}\right)^{1/2}\,,
\end{align}
where $\varepsilon$ is an arbitrarily small parameter satisfying $\varepsilon\ll k$
so that the Gaussians are very sharp, $F(q)$ is an arbitrary function of $q$ and $\bm{q}_+$ and $\bm{q}_-$ are such that
\begin{align}\label{eq:qpm}
|\bm{q}_+|=|\bm{q}_-|=k\quad,\quad \bm{q}_++\bm{q}_-=\bm{k}\,.
\end{align}
We can write explicitly a pair of vectors $q_{\pm,i}$ that satisfy the requirements \eqref{eq:qpm}
 by introducing a unit vector $\hat{s_i}$ orthogonal to $k_i$, i.e. $\hat{s}^ik_i=0$. In this way, we arrive at
\begin{align}\label{eq:qpm2}
\frac{q_{\pm,i}}{k}=\frac{1}{2}\hat{k}_i\pm \frac{\sqrt{3}}{2}\hat{s_i}\,,
\end{align}
where $\hat{k}_i$ is the unit vector of $k_i$.
By plugging in Eqs.~\eqref{eq:TGMD} and \eqref{eq:qpm2} into the last term of Eq.~\eqref{eq:heomMD2} we find that
\begin{align}\label{eq:projectionTT}
h^G_{\bm{k},\lambda}=h^N_{\bm{k},\lambda}-\tfrac{3}{2}e_\lambda^{ij}(\bm{k})\hat{s_i}\hat{s_j}k^2F^2(k)
+O\left(\varepsilon^2/k^2\right)\,,
\end{align}
To derive Eq.~\eqref{eq:projectionTT} we used that $\bm{k}-\bm{q}-\bm{q}_\pm=\bm{q}_{\mp}-\bm{q}$ 
and evaluated the integrand at $\bm{q}=\bm{q}_{\pm}$. 
This is a good approximation as long as $F(q)$ is a smoothly varying function of $q$. 
Since $\varepsilon$ can be arbitrarily small and we are interested only on subhorizon scales, the approximation is accurate enough. 
We may express the polarization tensor in terms of a third orthogonal vector $\hat{t_i}$ such that $\hat{k}_i\hat{t}^i=\hat{s_i}\hat{t}^i=0$ 
as $\sqrt{2}e^{ij}_+=\hat{s_i}\hat{s_j}-\hat{t_i}\hat{t_j}$ and $\sqrt{2}e^{ij}_\times=\hat{s_i}\hat{t_j}+\hat{t_i}\hat{s_j}$, for example. 
In this case, only the $\lambda=+$ contribution to the gauge transformation \eqref{eq:projectionTT} is non-vanishing.
Then setting
\begin{align}
F^2(k)={\frac{2\sqrt{2}}{3k^2}h^N_{k,+}}\,,
\end{align}
we can make the $\lambda=+$ polarization of the induced tensor modes \eqref{eq:hMD} arbitrarily small. 
We can do the same for the $\lambda=\times$ polarization by simply rotating $\hat{\bm{s}}$ 
and $\hat{\bm{t}}$ 
by 45 degrees.
It should be noted that the resulting form of $T_G$ is well-behaved on short scales as its amplitude 
very crudely decays as $T_G(k)\sim {\Phi_N(k)}/{k}$. 
This implies that indeed the induced GWs generated in a dust dominated universe may be removed by a small change of gauge. 

\section{Conclusions \label{sec:conclusions}}

Induced GWs are a very important probe of the early universe and inflation. They are even more relevant in light of future GW detectors such as LISA \cite{Audley:2017drz}, Taiji \cite{Guo:2018npi}, Tianqin \cite{Luo:2015ght}, DECIGO \cite{Seto:2001qf,Yagi:2011wg}, 
AION/MAGIS \cite{Badurina:2019hst}, ET \cite{Maggiore:2019uih} and PTA \cite{Lentati:2015qwp,Bian:2020bps}. 
However, some doubts were expressed in the literature about the gauge invariance of the induced GW spectrum \cite{Hwang:2017oxa}. 
If that were the case, it would pose serious problems for the theoretical predictions of induced GWs. Proposed solutions so far rely on 
either finding gauge invariant
 variables \cite{Wang:2019zhj,DeLuca:2019ufz,Yuan:2019fwv,Nakamura:2019zbe,Chang:2020tji,Chang:2020iji,Chang:2020mky}
 or arguing which would be the most suitable gauge \cite{DeLuca:2019ufz,Inomata:2019yww}. 
While these studies go certainly in the right direction, they are not a satisfactory solution of the gauge dependence pointed out
 in Ref.~\cite{Hwang:2017oxa}. The main reasons are two-fold: ($i$) a particular choice of gauge invariant variables is very much like 
 choosing a particular gauge fixing procedure and ($ii$) one should be able to express the observable quantity in different gauges, 
 albeit physically meaningful ones, and find the same GW spectrum in an arbitrary cosmological background.

In this paper, we made a new attemp to clarify the issue of the seeming gauge dependence of the induced GW spectrum. 
We showed that:
\begin{list}{}{}
\item[($i$)] 
The GW spectrum is only meaningful for subhorizon modes when the source term is no longer active. 
This is in analogy with the GWs from the mergers of binary black holes, where the GWs are well-defined far enough from the source.
\item[($ii$)] 
The Newton (or shear-free) gauge is suitable for both the calculations and physical interpretations of the induced GWs. 
This is because the equations of motion for the tensor modes $h_{ij}$ at second order in a general gauge \eqref{eq:heom},
 naturally gives the simplest form of the source term in the Newton gauge \eqref{eq:heomN}, and because
 the Newton gauge is well-behaved on the smallest scales where the Newton gravity limit is recovered. 
\item[($iii$)] 
The GW spectrum is gauge invariant under a set of reasonable gauge transformations and independent of the cosmological background. 
We define the set of reasonable gauges as those gauges which are well-behaved deep inside the horizon. 
More concretely, we require that after a gauge transformation the resulting trace part of the metric perturbation is of the order of or smaller than
the gravitational potential in the Newton gauge \eqref{eq:condition1}. 
This sets certain conditions on the gauge parameters given by Eq.~\eqref{eq:condition2}.
\end{list}
We discussed the point $(iii)$ for three different cases in a universe dominated by a perfect fluid with a constant equation of state $w$ and
sound speed $c_s^2=w$. We found that well-behaved gauges on small scales are, for example, 
 the flat slicing, constant Hubble slicing and synchronous gauges. 
 However, special attention has to be paid to the residual gauge ambiguity in the case of synchronous gauge as it may lead to spurious GWs. 
 We also showed that the comoving slicing gauge violates the condition $(iii)$ and therefore the induced GW spectrum
 in the comoving slicing gauge differs from the one in the Newton gauge. We argued that the comoving slicing gauge is a poor choice of gauge 
 to describe subhorizon physics as it would be equivalent to describing a GW detector in an oscillating coordinate frame, which could 
 naturally lead to gauge artifacts.

It is important to note that the requirement ($i$) excludes the case of induced GWs in a universe dominated by dust,
 i.e. a perfect fluid with $w=c_s^2=0$. In this case,  the Newton potential is constant on all scales and, therefore, the source term for the induced GWs is 
 continuously active throughout the dust dominated stage.  This does not imply that our criteria ($i$) - ($iii$) are wrong. 
  First, as we argued in Sec.~\ref{sec:dustdom}  the tensor modes induced in a dust universe can be hardly regarded as GWs. 
 They do not behave as a radiation fluid in an expanding background and their time-independence renders them unnoticeable by GW detectors. 
 Second and more importantly, we showed in Sec.~\ref{sec:dustdom} that the induced GWs during a dust dominated universe 
 can be gauged away by a small change of gauge. This means that one must follow the induced GWs until the universe transitions to 
 another stage and is dominated by a fluid with $c_s^2\neq 0$. 
 In the subsequent stage the induced GW spectrum is well-defined and invariant under the set of gauge transformations of point $(iii)$. 
 Our conclusion is in line with Refs.~\cite{Inomata:2019zqy,Inomata:2019ivs}, where they show that the dominant contribution to
  the induced GWs  in a dust dominated universe is always generated right after reheating. 
  Thus, after the universe is reheated our points ($i$) - ($iii$) apply and the induced GWs are approximately gauge invariant. 

Although we have focused on the induced GWs in a universe with a constant equation of state, we expect that the same conclusions 
would hold in more general situations. Furthermore, a definitive solution to the gauge issue would be to find a definition of the effective
energy density of GWs which is invariant under general gauge transformations at second order.
Despite being a very interesting direction, it is our of the scope of this paper. We leave these issues for future work.

\section*{Acknowledgments} 
G.D. would like to thank J-O.~Gong, K.~Inomata, M.~Kamionkowski, S.~Matarrese and S.~Pi for insightful discussions. We would like to thank J.~Gurian, D.~Jeong, J-c.~Hwang and H.~Noh for useful comments. G.D. as a Fellini fellow was supported 
by the European Union’s Horizon 2020 research and innovation programme under the Marie Sk{\l}odowska-Curie grant 
agreement No 754496. M.S. was supported in part by the JSPS KAKENHI Nos.~19H01895, 20H04727 and 20H05853. Calculations of 
cosmological perturbation theory at second order were checked using the \texttt{xPand (xAct) Mathematica} package \cite{Pitrou:2013hga}.

\appendix

\section{Gauge transformation \label{app:gauge}}
In this appendix we review the gauge transformation of the perturbation variables due to an infinitesimal change of the coordinates, 
which is given by
\begin{align}
\bar x^{\mu}=x^{\mu}+\xi^{\mu}\quad,\quad \xi^{\mu}=(T,\partial^i L)\,.
\end{align}
In line with the main text we only focus on the scalar-type components. 
To be consistent with the Hamiltonian formalism in which this work is based we compute the gauge transformation 
using the exponential mapping of the Lie derivative \cite{Bruni:1996im,Nakamura:2003wk,Malik:2008im}.
 This means that a perturbation variable $Q$ changes under a coordinate transformation according to
\begin{align}\label{eq:exponentiallie}
\bar Q={e}^{{\cal L}_\xi}Q\,,
\end{align}
where ${\cal L}_\xi$ is the Lie derivative along $\xi^\mu$. 
With this recipe we find that at first order the perturbation variables under consideration in the barred coordinate frame read
\begin{align}\label{eq:1ordergauge}
\bar\alpha&= \alpha+T'+{\cal H}T\,,\\
\bar\beta&= \beta-T+L'\,,\\
\bar\phi&= \phi+{\cal H}T+\tfrac{1}{3}\Delta L\,,\\
  \bar E&= E+L\,,\\
  \bar{\delta\rho}&= \delta\rho+\rho'T\,,\\
  \bar v&= v-L'\,.\label{eq:1ordergaugeV}
\end{align}
Focusing only on the transverse-traceless component of the metric, 
we find that the tensor mode trasforms at second order as \cite{Domenech:2017ems}
\begin{align}\label{eq:2ordergauge}
\bar h_{ij}=h_{ij}-\widehat{TT}_{ij}\,^{ab}\Big\{&-2\partial_a\sigma\partial_b T+2\partial_a\partial_kL\partial_k\partial_{b}E\nonumber\\&+\partial_aT\partial_bT+\partial_a\partial_kL\partial_b\partial_kL\Big\}\,,
\end{align}
where $\widehat{TT}_{ij}\,^{ab}$ is the transverse-traceless projector which can be found, e.g., in Ref.~\cite{Domenech:2017ems}.

\section{Relation to commonly used variables \label{app:relation}}
In this appendix, we present the relations between the variables in the commonly used perturbative expansion of the metric such as in Refs.~\cite{Malik:2008im,DeLuca:2019ufz} and 
the variables used in this work. Usually, the metric is expanded as
\begin{align}\label{eq:confdecom3}
ds^2=a^2(\tau)\Big(&-N^2d\tau^2+2N_idx^idt\nonumber\\&
+\left(\delta_{ij}+2\widetilde\phi\delta_{ij}+2\partial_i\partial_j\widetilde E+\widetilde h_{ij}\right)dx^idx^j\Big)\,.
\end{align} 
Comparing with Eq.~\eqref{eq:confdecom} we see that the only difference with our work is the spatial components of the metric. 
Expanding and matching we find that at first order 
\begin{align}
\widetilde \phi=\phi-\tfrac{1}{3}\Delta E\quad,\quad \widetilde E=E\,,
\end{align}
and at second order
\begin{align}
\widetilde h_{ij}=h_{ij}+\widehat{TT}^{ab}_{ij}\left\{2\partial_a\partial^kE\partial_k\partial_bE+4\widetilde \phi\partial_a\partial_bE\right\}\,.
\end{align}
Note that we only focused on the first order matching of $\widetilde \phi$ and $\widetilde E$ since only the first order is necessary for the discussions in the main text. These are the main relations that are relevant to relate the main equations in this paper with those in the usual perturbative expansion. 

In terms of the expansion \eqref{eq:confdecom3}, the equations of motion for the secondary GWs read
\begin{align}\label{eq:heom2}
(\hat D_\tau-\Delta) \widetilde h_{ij}=\widehat{TT}^{ab}_{ij}\widetilde S_{ab}\,,
\end{align}
where
\begin{align}\label{eq:heomsource2}
\widetilde S_{ab}=&4\partial_a\Phi\partial_b\Phi+2a^2\left(\rho+p\right)\partial_aV\partial_bV\nonumber\\&+(\hat D_\tau-\Delta)\left[\partial_a\partial^k\widetilde E\partial_k\partial_b \widetilde E+4\widetilde \phi\partial_a\partial_b\widetilde E-\partial_a\sigma\partial_b\sigma\right]\,,
\end{align}
and we defined
\begin{align}\label{eq:definitions2}
\Phi&\equiv\widetilde\phi+\sigma\quad,\quad V\equiv v+\widetilde E'\,,\\
\sigma&\equiv\beta-\widetilde E'\,.
\end{align}
The terms in Eq.~\eqref{eq:heomsource2} that contain the variable $\widetilde E$ have been neglected in Refs.~\cite{DeLuca:2019ufz,Inomata:2019yww}. However, in the Newtonian gauge where $\sigma=\widetilde E=0$ both expressions \eqref{eq:heomsource} and \eqref{eq:heomsource2} agree. We also spell out the gauge transformation of the tensor modes at second order for completeness, which is given by
\begin{align}\label{eq:2ordergauge2}
\widetilde{\bar h}_{ij}=\widetilde h_{ij}+&\widehat{TT}_{ij}\,^{ab}\Big\{2\partial_a\sigma\partial_b T+2\partial_a\partial_kL\partial_k\partial_{b}\widetilde E\nonumber\\&-\partial_aT\partial_bT+\partial_a\partial_kL\partial_b\partial_kL+4{\widetilde \phi}\partial_{a}\partial_{b} L\nonumber\\&
+4{\cal H}T\partial_{a}\partial_{b}\widetilde E+4{\cal H}T\partial_{a}\partial_{b}L\Big\}\,.
\end{align} 

\section{Equations of motion and solutions in Newton gauge  \label{app:eom}}
In this appendix we recapitulate the equations of motion for the first order perturbation in the Newton gauge on a flat FLRW background. 
The equation of motion of first order variables in a general gauge can be found e.g. in Ref.~\cite{Gong:2019mui}.
 We also provide the basic formulas for the induced GWs.

The Friedmann equations and the energy conservation equation for the background are given by
\begin{align}
3{\cal H}^2&=a^2\rho\,,\\
{\cal H}^2+2{\cal H}'&=-a^2p\,,\\
\rho'+3{\cal H}(\rho+p)&=0\,.
\end{align}
The solution to the scale factor for a general equation of state $w$ is given by
\begin{align}
a(\tau)=a_0\left(\frac{\tau}{\tau_0}\right)^{-1-b}\quad,\quad
b\equiv\frac{1-3w}{1+3w}\,,
\end{align}
and $\tau_0$ is an arbitrary reference time. Note that we assume $w>-1/3$, that is, a decelerating universe. This implies $-1<b<\infty$ for $\infty>w>-1/3$.

At linear order, we have $u_{0,N}=-a\left(1+\Phi_N\right)$, $u_{i,N}=a\partial_iV_N$ in the Newton gauge,
and the Einstein equations read
\begin{align}
6{\cal H}\Phi_N'+6{\cal H}^2\Phi_N-2\Delta\Phi_N&=a^2\delta\rho_N\,,\label{eq:deltarho}\\
\Phi_N'+{\cal H}\Phi_N&=\tfrac{1}{2}a^2(\rho+p) V_N\,,\\
\Phi_N''+3{\cal H}\Phi_N'+\left({\cal H}^2+2{\cal H}'\right)\Phi_N&=-\tfrac{1}{2}a^2\delta p_N\label{eq:deltap}\,.
\end{align}
where we have used $\alpha_N=-\Phi_N$.

For an adiabatic perfect fluid with a constant equation of state parameter $w$, the curvature perturbation in the 
Newton gauge with the initial condition, $\Phi\to const.$ for $\tau\to0$, is given by
\begin{align}\label{eq:gravitationalpotential}
\Phi(x)=\Phi_{\rm p}(k)\,2^{b+3/2}\Gamma[b+5/2]\, (c_sx)^{-b-3/2} J_{b+3/2}(c_sx)\,,
\end{align}
where $\Phi_{\rm p}(k)$ is the primordial value on superhorizon scales, 
$J_b(x)$ is the Bessel function of the first kind of order $b$, and we have defined $x\equiv k\tau$.

At second order in perturbation theory one finds a very compact solution for the spectrum of induced GWs Eq.~\eqref{eq:heomN}, 
which is given by
\begin{align}
\overline{{\cal P}^N_{h}(k,\tau)}=
&8\int_0^\infty  dv\int_{|1-v|}^{1+v}du\left[\frac{4v^2-(1+v^2-u^2)^2}{4uv}\right]^2
\nonumber\\
&\times{\cal P}_{\Phi_N}(kv){\cal P}_{\Phi_N}(ku)\overline{I^2}(x,u,v)\,,
\label{eq:powerspectrum1}
\\
I(x,u,v)=&\int_{0}^x d\tilde x \,G(x,\tilde x) f(c_s\tilde x, u, v)\,,
  \label{eq:kernel1}
\end{align}
where ${\cal P}_{\Phi_N}$ is the primordial dimensionless power spectrum of $\Phi_N$ and
the overline in Eq.~\eqref{eq:powerspectrum1} denotes oscillation average. 
For the case at hand, using Eq.~\eqref{eq:gravitationalpotential}, the source $f$ in the kernel \eqref{eq:kernel1} is expressed as
\begin{align}
f(c_s\tilde x, u, v)&=\frac{2^{2 b+3} (uvc_s^2\tilde x^2)^{-b-{1}/{2}} \Gamma^2\left[b+{5}/{2}\right]}{(2+b) (3+2 b)}
\nonumber\\
\times&\Big(J_{b+{1}/{2}}(uc_s\tilde x) J_{b+{1}/{2}}(v c_s\tilde x)
\nonumber\\
+&\frac{2+b}{1+b}J_{b+{5}/{2}}(u c_s\tilde x) J_{b+{5}/{2}}(v c_s\tilde x)\Big)\,,
\end{align}
and the retarded Green function, after using the first order solutions to the tensor modes \cite{Domenech:2019quo}, reads
\begin{align}\label{eq:green2}
G&(x,\tilde x)=\frac{\pi}{2}\frac{\tilde x^{3/2+b}}{x^{1/2+b}}\theta(x-\tilde x)
\nonumber\\
&\times\left(J_{b+1/2}(\tilde x)Y_{b+1/2}(x)-Y_{b+1/2}(\tilde x) J_{b+1/2}(x)\right)\,.
\end{align}
Most important for the discussion in the main text is the time dependence of the kernel on subhorizon scales ($x\gg1$),
which is approximately given by
\begin{align}\label{eq:Igreen2}
I(x\gg 1,u,v)\propto x^{-1-b}\,.
\end{align}
This implies that the induced GW spectrum decays as 
\begin{align}
\overline{{\cal P}^N_{h}(k,\tau)}\propto x^{-2-2b}\,.
\end{align}
The general expression after integration of the kernel can be found in Ref.~\cite{Domenech:2019quo}.

\section{From Newton to other gauges  \label{app:transformationrules}}
In this appendix, we present the formulas to go from the Newton gauge to some of the commonly used gauges. 
This is used in the main text to compare the amount of induced GWs in different gauges. 
We consider first a change in the time coordinate and later on the synchronous gauge which requires a change in the time and spatial coordinates. 
We are mainly interested in the gauge transformations at late times (or for subhorizon scales), i.e. $k\gg{\cal H}$ or $x\gg1$. 
At early times (or for superhorizon scales), i.e. $k\ll{\cal H}$ or $x\ll1$, it is checked that the results in all gauges coincide with each other.

\subsection{Temporal gauge parameters} 

Focusing on a temporal gauge transformation, we set $L=0$ in Eqs.~\eqref{eq:1ordergauge} $-$ \eqref{eq:1ordergaugeV}. 
Then for the gauges of our interest, the temporal gauge parameters in terms of the newtonian potential $\Phi_N$ are respectively
found as follows.

{\it a. Comoving slicing:} 
A useful gauge for superhorizon scales is the so-called comoving slicing, where $v+\beta=0$.
The temporal gauge parameter from the Newton gauge to the comoving slicing gauge is given by
\begin{align}
T_C=V_N=\frac{2}{a^2(\rho+p)}\left(\Phi_N'+{\cal H}\Phi_N\right)\,,
\end{align}
where $C$ stands for ``comoving''.

{\it b. Flat slicing:}
 In this gauge the spatial curvature is homogeneous, $\phi-\Delta E/3=0$.
The temporal gauge parameter from the Newton gauge to the flat slicing gauge is given by
\begin{align}
T_F=-\Phi_N/{\cal H}\,,
\end{align}
where $F$ stands for ``flat''.

{\it c. Uniform Hubble slicing:} 
In this gauge the trace of the extrinsic curvature is homogeneous. In our parametrization we have that
\begin{align}
K\equiv\nabla_\mu n^\mu=3n^\mu\partial_\mu(\phi+\ln a)-\tfrac{1}{N}D_iD^i\beta\,,
\end{align}
where $\nabla_\mu $ is the covariant derivative of  the spacetime metric, $D_i$ is the covariant derivative with respect to the spatial metric,
and $n^\mu$ is the unit hypersurface orthogonal 4-vector which defines our slicing of spacetime given by
\begin{align}
n^\mu=\frac{1}{N}\left(1,-N^i\right)\,.
\end{align}
Thus we have
\begin{align}
\delta K^{(1)}=3\phi'-3{\cal H}\alpha-\Delta\beta\,.
\end{align}
The temporal gauge parameter from the Newton gauge to the uniform Hubble slicing gauge in Fourier space is given by
\begin{align}
T_H=\frac{3\Phi'_N+3{\cal H}\Phi_N}{3{\cal H}^2-3{\cal H}'+k^2}\,,
\end{align}
where $H$ stands for ``uniform Hubble''.

Given the temporal gauge parameters as above, it is straightforward to spell out their
asymptotic behaviors at late times (or in the small scale limit) where we have
\begin{align}\label{eq:gravitationalpotential22}
\Phi(x\gg 1)\approx  \Phi(k)\, (c_sx)^{-2-b}\cos\left(\frac{b\pi}{2}-c_sx\right)\,.
\end{align}
This gives
\begin{align}
T_C(x\gg1)&\approx\frac{\Phi(k)}{(b+2)(1+b)c_sk}(c_sx)^{-b}\sin\left(\frac{b\pi}{2}-c_sx\right)\,,
\\
T_F(x\gg1)&\approx-\frac{\Phi(k)}{(1+b)c_sk}\, (c_sx)^{-1-b}\cos\left(\frac{b\pi}{2}-c_sx\right)\,,
\\
T_H(x\gg1)&\approx\frac{3c_s\Phi(k)}{k}(c_sx)^{-2-b}\sin\left(\frac{b\pi}{2}-c_sx\right)\,.
\end{align}

For completeness we show their respective superhorizon behavior, that is when $x\ll1$. We obtain
\begin{align}\label{eq:gravitationalpotential3}
\Phi(x\ll 1)\approx  \Phi_p(k)\,,
\end{align}
where $\Phi_p(k)$ is the initial conditions set by the primordial spectrum. The relation between $\Phi_p(k)$ and $\Phi(k)$ in Eq.~\eqref{eq:gravitationalpotential22} is given by
\begin{align}
\Phi(k)=-\Phi_p(k)\frac{2^{b+2} \Gamma\left[b+\frac{5}{2}\right]}{\sqrt{\pi }}\,.
\end{align}

Then, we find that the gauge parameters in the superhorizon limit read
\begin{align}
T_C(x\ll1)\approx\frac{2}{3}T_H&(x\ll1)\approx\nonumber\\& \frac{2\Phi_p(k)}{3{\cal H}}\frac{2 \Gamma\left[b+\frac{5}{2}\right]}{(2 b+3) \Gamma\left[b+\frac{3}{2}\right]}\,,
\\
T_F(x\ll1)\approx -\Phi_p&(k)/{\cal H}\,.
\end{align}
\subsection{Synchronous gauge} 
We treat the synchronous gauge separately as it is a subtle choice of gauge. The synchronous gauge corresponds to $\alpha=0$ and $\beta=0$.
It is defined by a geodesic congruence. However, as there is plenty of  freedom in the choice of a geodesic congruence,
the synchronous gauge is not fully specified by the gauge conditions $\alpha=\beta=0$.
Although this should not be a problem for the observable quantities, it is somewhat delicate in the mathematical formulation. 

The subtlety is made clear once we compute the temporal and spatial gauge parameters from the Newton gauge 
to the synchronous gauge. We find that in Fourier space
\begin{align}
T_S&=\frac{1}{a(\tau)}\int_0^\tau d\tau_1 a(\tau_1)\Phi_N(\tau_1)+\frac{T_0(k)}{a(\tau)}\,,\\
L_S&=\int_0^\tau d\tau_1T_S(\tau_1)+L_0(k)\,,
\end{align}
where $T_0(k)$ and $L_0(k)$ are arbitrary functions of the wavenumber $k$ which are constant in time. 
This means that the relation between the Newton gauge and the synchronous gauge is defined up to arbitrary $T_0(k)$ and $L_0(k)$. 
This gauge ambiguity proves crucial in the calculation of gravitational waves.

Here let us spell out the asymptotic behaviors of $T_S$ and $L_S$ at late times. 
\begin{widetext}
For the temporal parameter we find
\begin{align}\label{eq:TSAp}
T_S(x\gg1)&\approx\frac{(c_sx)^{-(1+b)}}{c_sk}\left(-{\Phi(k)}\frac{\sqrt{\pi}\Gamma[1+b/2]}{2\Gamma[(3+b)/2]}+ \frac{c_skT_0(k)}{a_0 (c_sx_0)^{-(1+b)}}\right)
-\frac{(c_sx)^{-(2+b)}}{c_sk} {\Phi(k)}\sin\left(\frac{b\pi}{2}-c_sx\right)\,.
\end{align}
For the spatial parameter we have
\begin{align}\label{eq:LSAPP}
L_S(x\gg1)\approx \frac{(c_sx)^{-b}}{ c_s^2k^2}
\left({\Phi(k)}\frac{\sqrt{\pi}\Gamma[b/2]}{4\Gamma[(3+b)/2]}-\frac{c_skT_0(k)}{ba_0 (c_sx_0)^{-(1+b)}}\right)&-\frac{2^{-1-b}\sqrt{\pi}}{b\Gamma[b+3/2]}\Phi(k)
\nonumber\\
&+L_0(k)-\frac{(c_sx)^{-(2+ b)}}{c_s^2k^2}{\Phi(k)} \cos\left(\frac{b\pi}{2}-c_sx\right)\,,
\end{align}
\end{widetext}
where the subscript $0$ in $a_0$ refers to evaluation of $a$ at some pivot time $\tau_0$, and we have defined
\begin{align}
x_0\equiv k\tau_0\,.
\end{align}
It should be noted that the case $b=0$ corresponds to a radiation dominated universe with $w=c_s^2=1/3$. 
Although there are terms that diverge in the limit $b\to0$ in Eq.~\eqref{eq:LSAPP}, it does not cause a problem.
We only need to displace the lower limit of the integral in Eq.~\eqref{eq:LSAPP}
to a non-zero $\tau$, which gives rise to $\log x$ terms after integration. 

Let us explicitly write for completeness the superhorizon behavior of the gauge parameters. We obtain
\begin{align}\label{eq:TSAp2}
T_S(x\ll1)&\approx \frac{T_0 x^{-(1+b)}}{a_0x_0^{-(1+b)}}+\Phi_p(k)\frac{x}{(2+b)k}\,,\\
L_S(x\ll1)&\approx L_0-\frac{T_0 x^{-b}}{bka_0x_0^{-(1+b)}}+\Phi_p(k)\frac{x^2}{2k^2(2+b)}\,.
\end{align}
It should be noted that the requirement that $T_S$ and $L_S$ are well-behaved on superhozion or subhorizon scales gives different values for the residual gauge parameters $T_0$ and $L_0$. The former sets $L_0=T_0=0$ while the latter from Eq.~\eqref{eq:condition2} requires
\begin{align}\label{eq:T0L0sub}
T_0&=a_0 (c_sx_0)^{-(1+b)}{\Phi(k)}\frac{\sqrt{\pi}\Gamma[1+b/2]}{2c_sk\Gamma[(3+b)/2]}\\
L_0&=\frac{2^{-1-b}\sqrt{\pi}}{b\Gamma[b+3/2]}\Phi(k)\,.
\end{align}

\begin{center}
\vspace*{3mm}
\noindent\rule[0.5ex]{0.4\linewidth}{0.6pt}
\end{center}

Let us emphasize that one may do any gauge transformation within the synchronous gauge at any time to achieve either well-behaved superhozion or subhorizon physics. For example, to solve the equations of motion \eqref{eq:heom} in the synchronous gauge the requirement of well-behaved $\sigma$ and $E$ on superhorizon scales is more appropriate, so that one may properly set the initial conditions. Contrariwise, to compute the induced GWs on subhorizon scales it is more reasonable to ask for a well-behaved $E$ on subhorizon scales.

\medskip

\bibliography{bibliofull.bib} 

\end{document}